\documentclass[prb,twocolumn,showpacs,preprintnumbers,amsmath,amssymb]{revtex4}    
\usepackage{graphicx}
\usepackage{dcolumn}
\usepackage{bm}

\preprint{submitted to RPB}

\begin{document}

\title{Surface Half-Metallicity of CrAs in the Zinc-Blende Structure}   

\author{I. Galanakis}\email{I.Galanakis@fz-juelich.de}

\affiliation{Institut f\"ur Festk\"orperforschung, Forschungszentrum J\"ulich, 
D-52425 J\"ulich, Germany}

\date{\today}

\begin{abstract}
The development of new techniques such as the molecular beam epitaxy have enabled 
the growth of thin films of materials presenting novel properties. Recently it was 
made possible to grow a CrAs thin-film in the zinc-blende structure. 
In this contribution, the full-potential screened KKR method is used to study the 
electronic and magnetic properties of bulk CrAs in this novel phase as well as 
the Cr and As terminated (001) surfaces. Bulk CrAs is found to be a
half-metallic ferromagnet for all three GaAs, AlAs and InAs experimental lattice 
constants with a total spin magnetic moment of 3 $\mu_B$.
The Cr-terminated surface retains the half-metallic character of 
the bulk, while in the case of the As-termination the surface states destroy the gap 
in the minority-spin band. 
\end{abstract}

\maketitle

The mainstream charged electronics has ignored the electron spin but recently a new
field, the so called magnetoelectronics or spintronics, has emerged that combines
magnetic elements with the existing conventional electronics to 
produce devices with new or enhanced properties.\cite{Review}
The interest in this field has increased after the simultaneous discovery of 
giant magneto-resistance (GMR) by the groups of Fert\cite{Fert88} and 
Gr\"unberg.\cite{Grunberg88} 
Although the achieved progress,
there are still central problems that have not been well solved. The injection
of spin polarized current from a ferromagnet into a semiconductor remains still an
open challenge.\cite{Datta90}  The most successful attempts concern the injection
of spin-current from a dilute magnetic semiconductor like GaMnAs, where Mn atoms 
have substituted Ga atoms. Ohno \textit{et al.}\cite{Ohno99} and Fiederling 
\textit{et al.}\cite{Fiederling} have used such contacts to inject spin-polarized electrons and holes,
respectively, into GaAs obtaining an
efficiency of 90\% spin-polarized current in GaAs. 
The advantage of dilute magnetic semiconductors is their 
coherent growth on semiconductors and their half-metallicity, 
\textit{i.e.} the minority band presents a
gap at the Fermi level, and thus electrons at the Fermi level are 100\%\ 
spin-polarized.\cite{Akai98} But existing magnetic semiconductors have very low Curie 
temperature, $T_c$, and thus are unattractive for industrial applications.
Other known half-metallic materials are CrO$_2$ and 
La$_{0.7}$Sr$_{0.3}$MnO$_3$,\cite{Soulen98} thin films of which have been found 
to present practically 100\%\ spin-polarization at the Fermi level at low 
temperatures\cite{Soulen98,Park98} but they also have low $T_c$.
Finally the half-metallic Heusler alloys like
NiMnSb\cite{Groot83,iosif1} present a $T_c$ far above the room temperature, 
but their surfaces are not half-metallic\cite{Groot01} and 
experimentally it is  difficult to control the stoichiometry of their 
surfaces.\cite{Ristoiu00}

\begin{figure}
\includegraphics[scale=0.35]{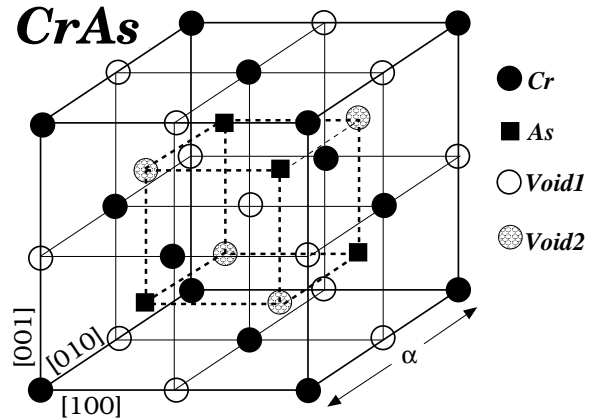}
 \caption{\label{fig1} Schematic representation of the zinc-blende structure. To model 
the system in our calculations we assume the existence of two non-equivalent 
vacant sites. The lattice consists of 4 fcc sublattices. The unit cell is that of an 
fcc lattice with four atoms per unit cell: Cr at $(0\:0\:0)$, As at $({1\over4}\:{1\over4}\:{1\over4})$
and the two vacant sites at  $({1\over2}\:{1\over2}\:{1\over2})$  and $({3\over4}\:{3\over4}\:{3\over4})$.}
\end{figure} 

Akinaga and collaborators have managed to grow thin films of CrAs on GaAs(100) 
substrates by molecular-beam epitaxy.\cite{Akinaga} They have found that CrAs is ferromagnetic
at room temperature with a $T_c$ larger than 400 K and they have deduced a 
total spin-magnetic moment of 
3 $\mu_B$. Bulk CrAs adopts either the MnP-type structure showing a 
helimagnetic-paramagnetic transition 
at 256 K, or it crystallizes as Cr$_2$As which is an antiferromagnet with a 
N\'eel temperature of 393 K. So the structure of the thin film cannot be one 
of the two stable bulk structures. Akinaga \textit{et al.}  have made the assumption 
that CrAs adopts the  zinc-blende (zb) structure of GaAs, presented in Fig.~\ref{fig1}, 
and using the full-potential linearized 
augmented-plane-wave  (FLAPW) method they have shown that in the ferromagnetic case 
zb-CrAs would be a half-metal with a total spin magnetic moment of 3 $\mu_B$ in agreement with
the experiment. Afterwards, Shirai\cite{Shirai01} continued the theoretical study of
the $3d$-transitional monoarsenides and showed that the ferromagnetic phase
for zb-CrAs should be more stable than the antiferromagnetic solution.
He also  calculated the theoretical equilibrium lattice constant and found a value
of 0.58nm, which lies inbetween the experimental lattice constants of GaAs (0.565nm), 
AlAs (0.566nm) and InAs (0.606nm). Due to the existence of 
two vacant sites per unit cell in the zb structure, 
CrAs can adopt the lattice constant of all GaAs, AlAs or InAs for the first few 
monolayers without deforming due to strain. So CrAs in the zinc-blende structure combines
the advantages of a very high $T_c$ with the coherent growth on GaAs or InAs.
In this contribution I  will study   
the electronic and magnetic properties of bulk CrAs
and its (001) surfaces taking into account both possible 
terminations, Cr or As, for different lattice constants   using an \textit{ab-initio} technique.
I  will show that the surface terminated
at Cr keeps the half-metallic character of the bulk CrAs, making it 
a serious candidate for an ``ideal'' spin-injection into GaAs, AlAs or InAs.

To perform the calculations, I  used the Vosko, Wilk and Nusair
parameterization\cite{Vosko} for the local density approximation
(LDA) to the exchange-correlation potential to solve
the Kohn-Sham equations within the full-potential screened Korringa-Kohn-Rostoker 
method.\cite{Papanikolaou}  
I have used the experimental lattice constants of GaAs and InAs for all the calculations;
I do not present the results for the AlAs lattice constant since the latter one 
is practically the same as the GaAs lattice parameter.
I  simulated the surface by a slab consisting of 15 CrAs layers, so that
I  have two equivalent surfaces.
For the screening I took for all compounds
interactions up to the sixth neighbors into account leading to a
tight-binding (TB) cluster around each atom of 65 neighbors. To
calculate the charge density, I  integrated along a contour on the
complex energy plane, which extends from the bottom of the band up
to the Fermi level,\cite{Zeller82} using 42 energy points. For the
Brillouin zone (BZ) integration I have used a 
\textbf{k}-space grid of 30$\times$30$\times$30 in the full BZ
for the bulk calculations and a \textbf{k}$_\parallel$-space grid 
30$\times$30 in the two-dimensional full BZ for the surface calculations. 
In addition I  used a cutoff  of $\ell_{max}$=6 for
the mutlipole expansion of the charge density and the potential
and a cutoff of $\ell_{max}$=3 for the wavefuctions. 

\begin{table}
\caption{\label{table1} Spin magnetic moment in $\mu_B$ for CrAs in the bulk case 
and for the surface and subsurface layers in the case of the Cr and As terminated
(001) surfaces for both the GaAs and InAs experimental 
lattice constants.}
\begin{ruledtabular}
\begin{tabular}{rlrrrrrr}
 \multicolumn{2}{c}{$m^\mathrm{spin}$($\mu_B$)} &  Cr   &    As  &  Void1  &  Void2 & Total\\ \hline
$a_\mathrm{GaAs}$ & Bulk      &  3.017   & -0.198 & 0.005 & 0.122 & 2.946 \\
       & (001)Cr &  3.961   & -0.177 & 0.078 & 0.170 &        \\
       & (001)As &  2.407   & -0.388 &-0.044 & 0.014 &\\
$a_\mathrm{InAs}$ & Bulk      &  3.269   & -0.382 &-0.029 & 0.080 & 2.937 \\
       & (001)Cr &  4.138   & -0.333 & 0.044 & 0.124 & \\
       & (001)As &  2.941   & -0.618 &-0.067 & 0.003 &
\end{tabular}
\end{ruledtabular}
\end{table}

\begin{figure}
\includegraphics[scale=0.42]{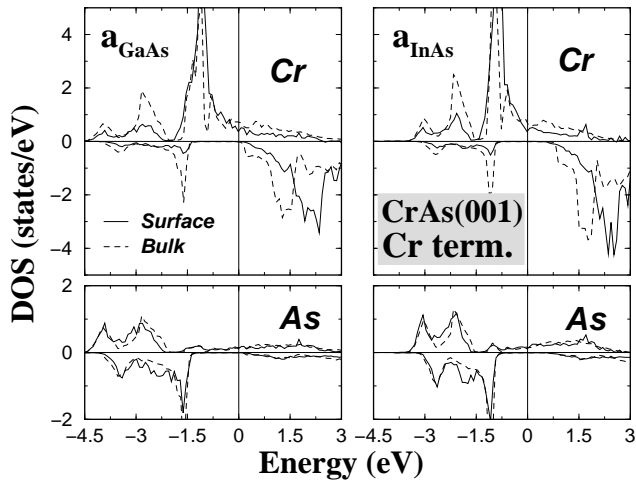}
 \caption{\label{fig3}Spin- and atom-resolved DOS of  the 
Cr-terminated (001) surface for the Cr atom at the surface layer and 
the As atom at the subsurface layer
for both the GaAs and InAs lattice constants. The As $s$ states are located 
at around 10 eV below the Fermi level and are not shown in the 
figures. The surface DOS are compared to the
bulk calculations (dashed lines).}
\end{figure}

In the zb structure there is a gap in the paramagnetic materials which 
is separating the bonding from the  antibonding states.
While in a semiconductor like GaAs only the bonding states are occupied 
and the semiconducting phase is stable, in the case of CrAs there are also  
occupied antibonding states. Due to the very high DOS at the Fermi level the paramagnetic 
phase is not stable and the ferromagnetic state is stabilized. This stabilization 
is not followed by a change in the charge transfer. For the GaAs lattice constant the 
Cr atom looses about 0.9  electrons and the As atom about 0.8 electrons for both the paramagnetic 
and  ferromagnetic phases and this charge is gained by the vacant sites.
The KKR calculated ferromagnetic DOS presented in  the left panel of Fig.~\ref{fig3} is similar to the 
one obtained using the FLAPW by Akinaga \textit{et al.}\cite{Akinaga} for the same 
lattice constant but  the Fermi level in the FLAPW  calculation was at the middle of
the gap while in my calculations it is at the right edge of the gap.  
In the ferromagnetic phase the Cr moments are well localized due to the 
exclusion of the spin-down electrons at the Cr site, similar to the localized Mn moments in the 
Heusler alloys,\cite{Kubler83} and the Cr spin moment is more than
3 $\mu_B$ as can be seen in Table \ref{table1}. The As atom possesses a small 
induced spin 
magnetic moment that is antiferromagnetically coupled to the Cr spin moment, while 
the contribution of the two vacant sites to the total moment is negligible. 

GaAs is not the only possible buffer for the growth of CrAs as InAs is also
widely used in experiments. In  Fig.~\ref{fig3} I have drawn the atomic and spin-projected
DOS for CrAs for the case of the experimental GaAs and InAs lattice constants (left
and right panel, respectively).
In both cases the system remains  half-metallic with a 
pretty large band gap (a width of $\sim$2 eV)  compared to the NiMnSb that
has a band gap that is only $\sim$ 0.5 eV wide.\cite{Groot83,iosif1} The $d$ states 
of Cr are well localized and in the case of the 
GaAs lattice constant it is mainly the $p$ states  of As which are squeezed compared to
the larger InAs lattice constant. As they are squeezed they move higher in energy pushing 
also the  the Fermi level higher in energy and this is clearly seen in Fig.~\ref{fig3}
where the Fermi level for the InAs lattice constant is at the middle of the gap while 
for the smaller GaAs lattice constant is is at the right edge of the gap.
The theoretical lattice
constant is inbetween these two values.\cite{Shirai01} So bulk CrAs is a 
stable half-metallic ferromagnet over a large range of lattice constants.

It is interesting also to mention that the total spin magnetic moment 
in a half-metallic material should be an integer number as the total 
number of spin-down occupied states is an integer number. As shown in Table \ref{table1}
the total spin-moment of bulk CrAs should be 3 $\mu_B$. In the KKR method the charge 
density is calculated by integrating on the complex energy plane. The numerical 
accuracy of this integration combined with the finite $\ell$-cutoff 
are responsible for the non-integer value of the total bulk spin moments presented 
in Table \ref{table1}.
The explanation of why there are exactly 3 uncompensated spin states is similar to the 
one for the half-metallic Heusler alloys which is discussed in detail in 
Ref.~\onlinecite{iosif2}.  The As atom in the
minority band offers one $s$ band and three $p$ bands that lie low in energy and can 
accommodate 4 electrons. There are practically no occupied minority $d$ states. 
The total number of uncompensated spins  should be just the total number of valence 
electrons, 11, minus two times the number of occupied minority states, 2$\times$4=8, 
and the total spin moment in $\mu_B$ would be also (11$-$8) $\mu_B$= 3 $\mu_B$. 
In the case of the InAs lattice constant the Cr
spin moment is larger than for the GaAs lattice constant. When   
the lattice is expanded, the Cr electronic structure becomes more atomiclike and 
its spin-magnetic moment increases. This is also reflected on the charge transfer which 
is smaller for the InAs lattice parameter (the Cr atom looses 0.8 electrons and the As 
atom 0.6). Although this means that the hybridization with 
As states should 
decrease, the larger Cr moment induces a larger spin-polarization of the As $p$
states, so that the total magnetic moment remains 3 $\mu_B$.

As the last part of this contribution
I  have studied the (001) surfaces of CrAs taking into account both 
the Cr- and As-terminations. I  present in Fig.~\ref{fig3} the atomic projected 
DOS for the Cr atom at the surface and the As atom at the subsurface layer for 
the Cr-terminated surface and for both the GaAs and InAs lattice constants.
There are no surface states
within the gap, so this surface is half-metallic. To my  knowledge this is 
the first case that electronic structure calculations predict a surface of an 
 intermetallic compound to present 100\%\ spin-polarization at the Fermi level. 
The As-atom  at the  subsurface layer has the same tetrahedral environment as in the 
bulk and the amount of electronic charge, which it looses, is similar to the bulk 
case.
The DOS of the surface atoms is slightly different than the bulk DOS.
The Cr atoms at the surface loose electronic charge towards the vacuum and their 
magnetic moment increases considerably by about $\sim$ 0.95 $\mu_B$ for both the 
GaAs and InAs lattice constants as  can be seen in Table \ref{table1}.
On the other hand the absolute value of the magnetic moment of the  As atom at the 
subsurface layer  decreases compared to the bulk case. Note that in agreement with the 
half-metallicity of this surface also the total spin moment remains an integer number. If the
total spin moments of the atoms in the surface and subsurface layers are added the total spin moment 
is slightly larger than 4 $\mu_B$, but if also the negative moment of the vacuum is added then the total moment is exactly
4 $\mu_B$ for both the GaAs  and InAs lattice constants.

\begin{figure}
\includegraphics[scale=0.42]{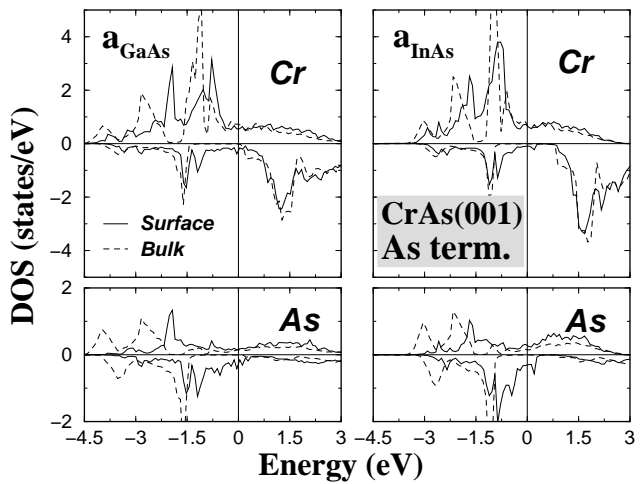}
 \caption{\label{fig4}Similar to Fig.~\protect{\ref{fig3}} for the As-terminated
(001) surface.}
\end{figure}

In the case of the As-terminated (001) surface the situation is not as ideal as in the 
case of the Cr terminated surface as can be seen in Fig.~\ref{fig4}. 
Now there are states 
within the gap that destroy the half-metallic character. In the case
of the InAs lattice constant the gap is only partially destroyed but the 
Fermi level is below the remaining fully spin-polarized region. Contrary to the 
Cr-termination the DOS of both the As surface atom and the Cr subsurface atom
present large deviations from the bulk case. The As atom at the surface looses $\sim$ 0.4
more electrons than in the bulk CrAs.
Its spin magnetic moment is practically doubled. Although the Cr atoms at the 
subsurface layer present a charge transfer comparable to the bulk calculations, the 
changes  in their DOS are  pronounced, the spin imbalance decreases and their
magnetic moment is 0.3-0.4 $\mu_B$ lower than in the bulk CrAs.

The different behavior of the two terminated surfaces is possibly arising from 
the different origin of surface states in the 3$d$ transition metals and 
the $sp$ elements. The As atoms have mainly $s$ and $p$ type electrons which can 
participate in directional bonding similar  to what happens in a semiconductor like GaAs. 
In a simple model we can assume that As atoms interact through 
directional bonds made up of some kind of $sp$ hybrids. When we open the As terminated 
surface an As atom at the surface looses 4 from its  12 nearest As neighbors (note that the As 
atoms sit on a fcc lattice) and so dangling bonds are created.
These dangling bonds create a 
surface band within the minority spin-down gap. In the case of a transition metal,
the surface states arise from $d$-type atomiclike states, which are pinned near the 
Fermi  level acting like a virtual bound state.\cite{DesjSpan} 
These states, except in the case of the most 
close-packed surfaces, are not  enough broadened by the interaction between the surface 
atom and its neighbors and they form a surface band. In the case of Cr in 
CrAs the exchange splitting of its $d$ states is very large  
and thus  when we open the Cr-terminated surface there is no $d$-like atomic state 
that can be found within the minority gap and the surface remains half-metallic. 
Contrary to Cr,
the Ni terminated (001) surface in the case of the NiMnSb compound looses its 
half-metallicity\cite{Groot01} because both the gap and the 
exchange splitting of the Ni $d$ states are  very small and when the 
Ni surface is opened there is a $d$-like state of the Ni atom at the
surface that is slightly shifted in energy compared to the continuum of the $d$ 
states and is pinned at the Fermi level.

I  have shown using first-principles calculations that the Cr-terminated 
(001) surface of CrAs in the zinc-blend structure should be half-metallic
for both the GaAs and InAs experimental lattice constants. Contrary to the 
Cr-terminated, the As-terminated surface looses its half-metallicity 
due to surface states within the gap. The bulk CrAs shows a total spin magnetic moment 
of 3 $\mu_B$ and its 
properties can be explained similarly to the ones of the half-metallic Heusler alloys.

\begin{acknowledgments}
The author acknowledges financial support from
the RT Network of {\em Computational
Magnetoelectronics} (contract RTN1-1999-00145) of the European Commission.
\end{acknowledgments}

\end{document}